# Unveiling Technorelief: Enhancing Neurodiverse Collaboration with Media Capabilities[1]

Maylis Saigot
Copenhagen Business School
msa.digi@cbs.dk

**Abstract**

*As the workforce settles into flexible work arrangements, researchers have focused on the collaborative and psychological consequences of the shift. While nearly a fifth of the world's population is estimated to be neurodivergent, the implications of remote collaboration on the cognitive, sensory, and socio-affective experiences of autistic workers are poorly understood. Prior literature suggests that information and communication technologies (ICTs) introduce major psychological stressors. Theoretically, these stressors ought to be exceptionally straining considering autistic traits – yet, studies describe a strong attraction to ICTs. We thus ask: how do digital technologies alleviate autistic workers' experiences of their collaborative work environment? Thirty-three interviews were conducted to address this question. Findings suggest that digital media present capabilities that filter input from the environment, turning it into a virtual stage that lets workers "time out". The resulting "technorelief" enables autistic workers to tune into their perceptions and regain control of their collaborative experiences.*

## 1. Introduction

In recent times, "neurodiversity" (Corker & French, 1999) has gained mainstream recognition, covering conditions like autism, ADHD, and dyslexia (Rosqvist et al., 2020). It challenges conventional views, celebrating natural neurological variations (Doyle, 2020). "Neurotypical" denotes typical brain function, while "neurodivergent" refers to different neurodevelopment (Jurgens, 2020). An estimated 15-20% of the world's population exhibit neurodivergence (NCI, 2022), often undiagnosed (cf. Malik-Soni et al., 2022), affecting teamwork with potentially 1 in 5 members being neurodivergent – knowingly or not. Given the acute variance within neurodiversity, this study focuses on autistic traits and contingencies, thus using autism as a revelatory case of neurodiversity (Yin, 2009).

Autism is a lifelong neurodevelopmental condition characterized by differences in communication, social interaction, and constricted and repetitive patterns of behavior, interests, or activities. It can be challenging for autistic adults to cope with the everyday triggers of ill-fitted work environments (Cooper et al., 2017). Autistic workers often value fixed schedules and structured interactions due to behavioral rigidity, a diagnostic criterion for autism (Petrolini et al., 2023). However, digital media's disruptions, like interruptions and work-life imbalance, conflict with these preferences (Ayyagari et al., 2011). Paradoxically, studies show autistic individuals are drawn to ICTs for social connection (e.g., McGhee Hassrick et al., 2021). Emerging research indicates that digital media hold promise for inclusive workplaces and emotion regulation (e.g., Zolyomi et al., 2019).

Thus, we ask: *how do digital technologies alleviate autistic workers' experience of their collaborative work environment?* This article sets out to re-actualize findings from previous research in light of the changing norms and expectations of post-pandemic work arrangements. Moreover, we seek to uncover how technology may benefit the cognitive, sensory, and affective mechanisms that underlie neurodiverse collaboration. To this end, we engaged in conversations with advocates and conducted semi-structured interviews with autistic participants. In the next section, we provide a short account of prior work to serve as background for the study. We then describe our research approach, including details on data collection and analysis. Next, we report our findings from the interview data. Finally, we discuss those findings in light of the current state of knowledge and develop a model of *technorelief*.

---

## 2. Prior Work

### 2.1 Masking

Autistic individuals typically engage in behaviors and strategies to fit into a neurotypical community more seamlessly (Hull et al., 2017). This set of complicated coping behaviors is often referred to as *camouflaging*, *masking*, or *compensation* (Pearson & Rose, 2021) and may include concealing atypical traits and using techniques to appear socially competent (Hull et al., 2017). Research suggests the main motivations for masking are assimilation (i.e., the pressure to blend in and "seem normal" to attain social and professional goals and protect their safety and well-being) and connection (i.e., the desire to overcome initial obstacles to connection, allowing to develop future relationships) (Hull et al., 2017). Neurotypicals shift identities based on context (Scheepers & Ellemers, 2019), whereas masking entails rejecting one's true self, leading to intense psychological stress (Pearson & Rose, 2021). For example, masking is associated with mental exhaustion, threats to self-perception, and suicidality (Hull et al., 2017; Pearson & Rose, 2021).

### 2.2 Technostress

Technostress refers to the stress experienced by individuals as a result of their use of ICTs (Ragu-Nathan et al., 2008). Technostress creators include factors that create stress from the use of ICTs (e.g., work and emotional overload, role ambiguity, mobbing, obstacles hindering ICT use, etc.). Technostress inhibitors include organizational mechanisms that reduce stress from the use of ICTs (e.g., organizational and technical support, end-user involvement in implementation, etc.). Stressors and inhibitors jointly impact job satisfaction, organizational commitment, and continuance. Technostress can substantially affect well-being: technostrain refers to feelings of anxiety, fatigue, skepticism, and inefficacy beliefs related to the use of technologies (Salanova et al., 2013). Research indicates individual traits influencing technostress. Higher neuroticism is linked to more technostress (Wang et al., 2020), as do external locus of control and low self-esteem (Korzynski et al., 2020; Zielonka & Rothlauf, 2022). Autistic traits often align with these factors, like higher neuroticism (Schriber et al., 2014; Schwartzman et al., 2016) and low self-esteem (Cooper et al., 2017). Common victimization (Trundle et al., 2022) and masking pressure reinforce an external locus of control (Pearson & Rose, 2021). Thus, autistic individuals appear particularly vulnerable to technostrain.

## 3. Research Approach

This study adopts a constructivist paradigm, embracing a relativist ontology and an interpretivist epistemology (Denzin & Lincoln, 2011). We aim to understand participant perspectives, unraveling the mechanisms behind diverse experiences (Reid et al., 2005). Designed as an inductive qualitative study, this work uses flexible non-standardized data collection techniques allowing adjustments and iterations.

### 3.1 Data collection

In preparation, four initial unstructured interviews involved multi-cultural, self-proclaimed experts engaged in advocacy via social media, non-profits, or social entrepreneurship. These conversations refined the study scope, crafted the interview guide, established ethical practices, and formed a recruitment strategy. Relevant seminars, webinars, and discussions were also attended by the researcher.

Participant recruitment was primarily conducted over LinkedIn, where potential participants were identified via their public social media activity, including information in their description and engagement with content related to neurodiversity. We purposefully selected participants who were (i) clinically or self-diagnosed with autism (Lewis, 2017) and (ii) engaged in collaborative work over digital media. This includes remote, flexible, hybrid, and office-based workers spanning large distances. While there were no exclusion criteria based on geographical location, the disproportionate Western representation in LinkedIn users made it challenging to recruit participants from other regions. Self-diagnosis was included to mirror real-life workforce diversity. A total of 29 semi-structured interviews were conducted with participants (see Table 1 for full details). Interviews, lasting 51 to 134 minutes, were conducted via video or audio calls, written interview sheets, or synchronous text messaging as preferred by participants. We prioritized informed consent and comfort, conducting quality checks and offering choices related to the use of the camera, virtual background, and closed-captioning. Language use (identity vs. person first) was tailored based on informant input.

**Table 1. Expert informants and study participants**

| Alias | Age | Disclosed condition(s) | Occupation | Gender | Location | Role |
|---|---|---|---|---|---|---|
| Cathy | 48 | ADHD, autism, SPD* | Entrepreneur** | Female | USA | Expert |
| Rebecca | 32 | Neurodivergent | Entrepreneur** | Female | Denmark | Expert |
| Sam | 50 | Autism | Manager | Male | USA | Expert |
| John | 52 | ADHD, autism | Entrepreneur** | Male | India | Expert |
| Anna | 33 | ADHD, autism, SPD* | Senior production manager | Non-binary | UK | Participant |
| Alexa | 57 | Autism, SPD* | UX design lead | Female | USA | Participant |
| Alice | 45 | Autism | Entrepreneur** | Female | USA | Participant |
| Amy | 37 | ADHD, autism | Product designer | Female | USA | Participant |
| Fred | 45 | ADHD, autism | Chief information security officer | Male | USA | Participant |
| Robbie | 37 | ADHD, autism | Quality laboratory technician | Agender | USA | Participant |
| Cathryn | 26 | Autism, depression, anxiety | Architectural assistant | Female | UK | Participant |
| Donald | 35 | Autism | Software engineer | Male | USA | Participant |
| Dennie | 46 | ADHD, autism, autoimmune disease | Software engineer, entrepreneur** | Male | UK | Participant |
| Ted | 54 | Autism | Data modeler | Male | UK | Participant |
| Margot | 35 | Autism | Lead business analyst | Female | UK | Participant |
| Bernard | 39 | Autism | Management consultant, Director** | Male | Australia | Participant |
| Alfred | 63 | Autism. SDP | Freelance software engineer | Male | UK | Participant |
| Ariel | 47 | Autism | Education and Disability Consultant | Female | USA | Participant |
| Isadora | 36 | ADHD, autism | Student (film), Entrepreneur | Female | USA | Participant |
| Jacob | 27 | ADHD, autism | Senior software developer | Male | UK | Participant |
| Jenny | 43 | Autism (non-speaking) | PhD Student**, Consultant** | Female | UK | Participant |
| Jeremy | 40 | Autism | Full stack web developer | Male | USA | Participant |
| Jonathan | 48 | ADHD, autism | Managing director (consultant)** | Male | UK | Participant |
| Anika | 44 | ADHD, autism | CEO and Founder** | Female | USA | Participant |
| Melanie | 32 | Autism | Senior learning designer | Female | USA | Participant |
| Nicole | 24 | Autism, connective tissue disorder | Coordinator for Policy & Outreach | Non-binary | USA | Participant |
| Polly | 55 | Autism | Education specialist**, entrepreneur** | Female | USA | Participant |
| Brittany | 48 | Autism | Senior disaster resilience officer | Female | Australia | Participant |
| Shawn | 26 | Autism | Chief technology officer | Male | USA | Participant |
| Tommy | 28 | Autism | Autism Advocate, Ph.D. Student | Male | Australia | Participant |
| Tania | 44 | Autism | Ph.D. Student, Research Assistant | Female | USA | Participant |
| Veronica | 34 | ADHD, autism, neuromuscular disease | Cyber quality assurance manager | Female | USA | Participant |
| Zelie | 31 | Autism, SDP | Financial services professional | Female | Ireland | Participant |

*Spatial Processing Disorder, **Neurodiversity-related activity*

### 3.2 Data analysis

Audiovisual interviews were transcribed using Konch.ai, supplemented by manual formatting and verification. Data organization, coding, and interpretation were managed with Atlas.ti. We followed a six-phase thematic analysis to identify and interpret patterns of meanings (Braun & Clarke, 2012). In Phase 1 (familiarization), recordings and transcripts were reviewed multiple times to get into an "insider's perspective". Subsequently, the researcher shifted to an interpretive stance, making sense of participants' perspectives (steps 2-6). (Reid et al., 2005). In phase 2 (initial coding), semantic and inductive coding resulted in broad categories of content (e.g., difficulties with social cues, advantages of remote work, etc.). In phase 3 (search of themes), latent codes were generated. Patterns became apparent as codes were merged or split. In phase 4 (reviewing themes), initial codes were reviewed and restructured into three emerging themes: (i) technology reduces exposure to work-related stimuli, (ii) the leaner sensory input makes it less straining to shape a response to the input, and (iii) the slower sensory input makes processing emotional and social cues more manageable. In phase 5 (theme naming and definition), themes were defined and interview quotes were mapped into them to validate their homogeneity (see Table 2). Phase 6 (report production) focused on ordering and organizing the themes into a coherent narrative while ensuring that all interpretations stayed true to the participants' words by paying attention to the broader context of the selected quotes. Follow-up questions were sent to participants

## 4. Findings

### 4.1 Digital filters

#### 4.1.1 Sensory input and cognitive overload

First, we find that in-situ work brings about multiple sensory challenges, with a preponderance of auditory processing issues. Participants often find it difficult to focus on a single voice, spatially identify where a sound is coming from, and determine whether someone is speaking to them or someone else (cf. Cathy, Anna, Alexa, Zelie). When such issues manifest publicly, they put the workers at risk of stigma – especially when others are unfamiliar with such disorders, their commonality, and their comorbidity with autism. In such cases, they can result in negative attributions or mockery, which challenges the employees' professional identity and self-esteem:

"*My manager used to say that I have like selective hearing*" (Zelie); "*[He] thought that was just really funny, so - not in a good way. So it's awkward sometimes*" (Anna).

Auditory processing issues also challenge people's ability to cope with their environment. It creates an additional cognitive burden, as they attempt to reconcile sometimes conflicting sensory input: "*People frequently walk up behind you and are talking to you, but they also walk up behind you and are talking to the people behind you. So you just don't know*" (Anna). This invisible labor can impede people's ability to engage in their daily tasks: "*That person is going to be exhausted and then they're not any good probably for the rest of the day*" (Cathy). Importantly, sensory stimuli can sometimes result in intense physical pain and emotional distress: "*There's […] a quality of sound plus loudness that actually can feel like a stabbing sensation*" (Anna). Characteristics of the space itself can also create additional challenges that capture mental resources, including the lack of control over the ambient temperature, distractions caused by people walking by, sensitivity to bright lights, and other types of physical discomfort: "*Let's just say the room's too hot or too cold or your clothes that day are uncomfortable, or the chair that you got stuck with is the worst one in the whole room and it squeaks*" (Cathy). Again, these uncontrollable variables cause distress and challenge the participants' ability to focus on work: "*There was constant movement all around the room, which again, I just can't help it, if there's movement I'm drawn to it. And it's very distracting*" (Ted). The combinational effects of sensory inputs often result in the temporary incapacity to engage with a situation, respond to a stimulus, control or regulate emotional responses: "*When I am overwhelmed, I would be tuning out or shutting down*" (Tommy). Most concerningly, participants described long-term cognitive, psychological, and physical impairments from repeated exposure to these environments: "*I would be having like digestive problems from like being in the noisy environments and like so drained as well*" (Zelie).

#### 4.1.2 The filtering effect of technology

Controlling the environment is a core benefit of remote work. Participants described several media capabilities that keep triggers at bay. First, being in their home environment allows them to control features of the room, such as ambient brightness, temperature, etc.: "*When I'm at home, I can put on more jumpers or I can put on heating or I can open the window*" (Zelie). Removing unnecessary distractions is paramount to achieving their work goals. Likewise, others' emotions can be distracting. Media capabilities like text formats help tone the intensity down, which helps some participants preserve their emotional stability. In this case, participants describe what seems to act as an emotional filter: "*I am emotionally very sensitive and can directly sense and feel other people's emotional states. It can be very intense. […] I read emails as they are less demanding cognitively and emotionally*" (Tommy). At the same time, dealing with other people's emotional outbursts takes away some of the finite attention that could otherwise be allocated to work tasks: "*I can sense and feel the emotional states of others, even online, which may make things even more challenging as I need to juggle the task requirements*" (Tommy). Much like other environmental factors, the text format helps reduce the incoming flow of stimuli as well as their negative consequences.

Physical office spaces are also characterized by important sets of unwritten rules. These rules are often taken for granted by neurotypical managers and employees but can be intimidating for others. For example, Jacob explains: "*There's a lot of unwritten rules that you might need to follow, whereas none of that exists online. It's all a lot more clear how things work*" (Jacob). As physical spaces are often perceived as chaotic, participants find that online environments make it easier to establish explicit social rules. For example, co-located colleagues might walk by their office and spark up a conversation. This behavior can annoy some participants as it interrupts their workflow and breaks their focus: *"[Spontaneous calls] used to happen a lot and that used to be really stressful [...]. So what I do now is ask people to put a time in my diary if they want to speak about something that's not relevant to what I'm currently working on"* (Jacob). The digital capability to communicate explicit rules by default when interacting with someone helps reduce and triage those requests: "I *don't like unsolicited communications, so I actually have on my email signature "Don't call me unless we've arranged it"*" (Bernard). This practice helps filter interaction to reduce the cognitive burden of task-switching and the emotional burden of social anxiety.

Furthermore, technology seems to be used both as a barrier and an enabler of social connection. Audiovisual channels help participants gain access to important social cues, such as tone or facial expressions while blocking out other elements of social interaction that could be challenging in person: *"It's much easier to catch [cues] on Zoom than it is in person. [...] Because I don't have to think about any physical logistics of space and how often the eye glance is or isn't [...]. You know, those kind of in-person things make it much more data-rich. And the more data-rich, dynamic, the easier it is to miss a microexpression"* (Polly). As such, audiovisual channels provide much more focused access to detailed facial features than in-person interaction does. This level of detail in a physical environment would come with ample background noise that would make it more challenging to catch subtle cues. Sometimes, the richness of the webcam feed is still too intense. In this case, the digital capability to access audio-only channels helps facilitate focused attention: *"I would have my head on the desk with my eyes closed with my head next to the telephone so that I could focus entirely on what they were saying and respond"* (Ted).

Another important digital capability is auto-generated live closed-captioning. Participants deplore that the systematic use of the feature is still limited, as it has a strong potential to significantly impact their ability to perform: "*But most people don't think about that. [...] If you expect me to take in information, process it, respond, engage in that moment...*" (Cathy). Other capabilities that help process fast-paced content include recording or transcribing an interaction. If a meeting is particularly fast or emotionally loaded, participants can use those capabilities to re-visit the content with greater focus and cognitive resources. It enables them to process relevant information without interference, which fosters contribution: "*So for me, the more time I have to digest information... I'm happy to attend the brainstorming meeting provided I do not have to contribute and that everything is written down and that I can then take that away with me. Digest it. Reflect on it, and then contribute later*" (Dennie).

Digital collaboration tools offer many capabilities that filter the incoming flow of sensory input and social triggers. They provide *à la carte* richness, by letting individuals pick and choose which layers of information to leave out or emphasize based on their needs and preferences, resulting in improved health and performance. Filtering capabilities can have tremendous positive effects on quality of life: "*And I'm kind of healthy and happy now and it's the first time I've been healthy and happy. It's really nice. I like it. I really do not want to go back to the office. The office environment is horrific*" (Ted).

### 4.2 The digital stage

#### 4.2.1 Masking to exhaustion

Most participants describe masking regularly. They often do it consciously and intentionally, for example, to de-escalate a conflictual situation or achieve socio-professional goals: "*I'll mask just to shut down a conversation*" (Alice). Masking is often described as a learned behavior, and seems to be carefully rehearsed and perfected, as described by Bernard: "*I aced the test for masking, and a lot of it comes down to some of my acting training*" (Bernard). However, as an active cognitive and behavioral process that is simultaneous with other ongoing activities, it is often exhausting: "*So it's exhausting to do all of that [...]. It absolutely drains more. And it's unnecessary*" (Polly). Importantly, masking can also have important consequences on mental and physical health. Participants explain it creates profound dissonance between their core identity and what they perceive to be an ideal identity to pursue. Even though they consider masking unhealthy, they often feel unsafe unmasking: "*That is not me being a super savvy, self-care person and that isn't being authentic either. That's masking. We mask all the time. It's really not safe to be authentic*" (Polly). For many participants who identify as "late-discovered", this awareness has come later in life, meaning that have lacked the support to engage in healthier coping mechanisms: "*What I thought I was doing was editing myself to be a better person*" (Anna). In some cases, the lack of knowledge and support has resulted in self-medication. Because masking leads to poor self-esteem and negative self-talk, participants became trapped in psychological cycles of sense-making infused with self-doubt. In those cases, the perceived hyperactivity of their mental processes made it difficult to restore emotional stability, seemingly leaving psychoactive substances as the most effective remedy. For example, Alfred used alcohol: "*I take the blame and it depresses me. This is the masking and then the neurodivergent thing. [...] Earlier, in earlier parts of my life, I used to drink an awful lot to knock myself out*" (Alfred).

### 4.2.2 The digital space as a controlled stage

When an actor steps onto a scene, cameras, lights, and props are all tailored to the act. As a result, the actor can dedicate their undivided attention to the act and feel confident that everything else is under control. Similarly, digital media provide capabilities to simulate a "digital stage" that is tailored to the interactive and social act, by filtering out many aspects of ambiguity, uncertainty, or unpredictability. This greatly reduces the strain of masking.

For example, eye contact is a core – yet exhausting – component of masking. During audiovisual interaction, participants use the webcam lens or LED as a proxy for their interlocutor's eye. This technique allows many participants to engage in typical eye contact behavior without experiencing the associated strain: "*I'm looking at the camera a lot because it's a nice little dot, and I can act like I'm looking in your eyes, but I'm looking at the camera. [...] In person it's really hard. [...] I will make myself look people in the eye because I was brought up to do that. But it's not easy*" (Anna).

Masking is also concerned with the dissimulation of behaviors that may be considered odd by neurotypical colleagues. However, this can be difficult when said behaviors are uncontrolled and often nonconscious. For example, Zelie finds it challenging to control her facial expressions, yet she was told she sometimes emotes inappropriately: "*I actually can't control my facial expressions at all. [...] So in that sense, [online] is great because, um, I can roll my eyes as much as I want and no one will see me, um, and I can control my voice better than my face*" (Zelie). The capability to select the channels she interacts through by controlling whether her webcam is on or off protects her from the negative attributions others often make based on her uncontrolled facial expressions. She can instead "pick her stage", as she feels more confident regulating her vocal tone than her facial expressions. Because emoting can be challenging, the digital capability to use virtual facial expressions (e.g., emojis) in text-based communication enables participants to effortlessly display emotions outwardly – which can otherwise be challenging during face-to-face interaction: "*Emojis on text, I use it quite a lot. [...] And I can get people by with me or that as well because they kind of, they get it, they get me. They tend to get me if I throw in a lot of emojis*" (Alfred).

Stimming behaviors are essential to participants' demeanor, and a critical sensory outlet. Their suppression can result in considerable distress: "*I almost started crying. [...] I couldn't hold it together that long. Because there was no outlet, you know, all this input, no output that I could do that I felt comfortable doing*" (Tania). Participants explain they feel more comfortable engaging in stimming when they can "hide it" from their interlocutors. Digital tools offer media capabilities to enact a natural interaction while concealing aspects that participants prefer to keep private. Specifically, the visual frame of the communication channel helps participants enhance their social performance because they do not have to worry about behaviors that fall outside of the curated frame: "*So like here, you know, you can only see me from here up, so you can't see that I'm fiddling with something or I'm shaking my leg or I'm moving my hands*" (Tania).

Being released from the perceived obligation to control various aspects of their conduct results in participants feeling more capable to perform their work well: "*So it's like all my resources are going to go to [my presentation]. I'm not going to worry about what I'm doing on camera*" (Tania). These additional cognitive resources can thus be allocated to other aspects of the interaction. For example, Anna explains that some of her most desirable personality traits can become more visible, which improves her experience working with colleagues: "*The thoughts are flying and, and there's no... I know why it's exhilarating. There's no monitoring of my face or my voice, you know?*" (Anna).

## 4.3 The digital timeout

### 4.3.1 Asynchronous socio-emotional processing

Participants described sometimes complex relationships with socio-affective and interoceptive cues. It can sometimes take additional time for participants to process a situation and experience the resulting emotion: "*You can have an emotional reaction at any amount of time after the event*" (Alfred). This can also happen due to the dynamism of a situation, which does not provide scope for emotional check-in: "*You see, with my sensory perceptual and information processing issues as well as additional communication needs, it is extremely difficult to be consciously aware of my emotions during an event*" (Tommy). These challenges can also be related to impaired: "*Since I experience high alexithymia and low interoception, I don't always feel these physical sensations. My thoughts get processed independently of my emotions, and I'm not sure emotions represent any objective reality*" (Robbie). Participants described needing extended periods before realizing that they are under intense stress: "*I do not notice emotions until they are big. This has had an effect in my work life in multiple ways, most often in the management of stress*" (Jenny). Dennie describes how not being in touch with his emotions promptly had major consequences on his relationships, especially in the context of work, as it translates into an "*explosive temper*".

Participants describe experiencing delayed processing of other people's emotions as well: "*I sense other people's energies quite strongly, which is one of many reasons I enjoy solitude. I don't always know what their emotions mean in real-time*" (Robbie). Oftentimes, dynamic interaction does not provide the time or space for participants to become fully aware of the emotions being communicated by their colleagues: "*When it comes to my attention that I've been misinterpreting someone's emotions, and I finally reach clarity, my mind resynchronizes all relevant memories into that context…*" (Veronica). In some cases, even when the emotional trigger is explicitly mentioned, the realization that the colleague may need emotional support does not come until after the interaction has taken place: "*I felt bad after it, like afterward about this situation because I did feel sorry for her. But it's like delayed*" (Zelie).

Catching and accurately interpreting socio-emotional cues can also be challenging, especially when it comes to receiving or providing feedback. In particular, participants experience intense distress when receiving negative feedback while being unable to estimate the gravity of the situation. As Donald puts it: "*I never really understood if my manager was angry or disappointed. Or just wanted me to improve or if I was going to get fired*". This is also true about missing positive cues. Jacob explains that he finds picking up on positive cues from others particularly challenging, which often results in negative attributions about what others think of him: "*I get very insecure about how people feel about me because I'm not picking up that they are expressing positive emotions about me when they are*" (Jacob). Struggling with interpreting cues from others can not only result in worse relationships and social well-being, but it can also breach the participants' self-esteem.

### 4.3.2 Digital timeout for controlled reflection

The filtering capabilities of digital media seem to create a parallel temporality, where participants can time out for self-reflection without interrupting the communication flow. In particular, participants value the digital capability to rehearse the content and form of a message.

Most participants exhibited strong work engagement and a desire to do a good job – even as some did not enjoy the job or feel close to their colleagues. These strong work ethics mean they often ask tough questions, challenge ways of working, and strive to achieve the best result: "*Sometimes people, for the sake of social niceties, not wanting to harm someone, will go along with an idea that they're not fully convinced of. Whereas I'm the other extreme*" (Jacob). These behaviors can be interpreted by colleagues and managers as a challenge to authority or pettiness, and sometimes result in people taking offense: "*But at that point, I feel like I'm being the one that's rude by trying to correct it because she's not listening*" (Tania). The ability to convey those messages in writing rather than during a live conversation can be advantageous because it makes it possible for participants to craft their message: "*I will write a message and then I'll read it and then I'll go, you know, ask them how they are, and hey, hope your Monday is going well or whatever*" (Anna). In that sense, the digital capability to rehearse is used to preserve others' feelings. While participants often describe not understanding their colleagues' emotions, many show deep care about their colleagues' emotional well-being. For example, Isadora describes a situation where she had to communicate negative feedback to a member of her team. Despite having rehearsed the (in-person) interaction, things got out of hand when her co-worker got emotional. She regretted not using text messaging, which would have made it easier to stick to the script and preserve her colleagues' feelings: "*And I was afraid to do that to her because, like, she doesn't deserve that*" (Isadora). While in-person interaction can be rehearsed, it seems that text-based communication makes it possible to rehearse and make sure the interaction follows the rehearsed path. Moreover, it also provides the psychological safety of knowing that the interaction will go as planned.

Media capabilities add a temporal buffer that also helps participants reflect on their thoughts and feeling before they respond to a communication: "*I find the option of a chat function especially helpful, enabling me to reflect on my words and reorder them as necessary before sending them into the conversation*" (Jenny). This timeout lets participants make an informed decision when they respond, either by being in tune with their inner reaction and knowing how to best communicate it outward – or by adopting a more socially acceptable response. Participants describe using artificial intelligence to leverage those digital timeouts, typing response drafts into an AI chatbot: "*Seeing an already written email will either help me phrase things that I'm already feeling, or I'll see something, know that it's wrong, and then want to express it in a different*" (Jacob). With this practice, they leverage the digital capability to generate communication prompts to produce a message that they feel is true to their thoughts and intentions. They may also paste messages from others that they find confusing, using the AI to help clarify subtle cues that may be concealed in wordiness. This is also helpful to tailor a more appropriate response: *"[Chat GPT] tells me I'm being insensitive*" (Jacob); "*I have turned to ChatGPT to soften and correct my content to get the same impact in more neutral words because my choices are historically bad at triggering a) neurotypicals b) those who are particularly narcissistic*" (Alfred).

Finally, media capabilities that instill more distance between participants and potentially triggering emotions from others make room for making sense of and rationalizing their emotional experiences: "*Well, I have more of a barrier, I suppose, from that emotion. It's easier for me to tell what's my emotion versus someone else's*" (Brittany). This is especially helpful when participants feel vulnerable to negative emotions and thus susceptible to being substantially emotionally affected: " *So that's why the ability to do remote meetings and things of that caliber, I can let go of that toxicity in an environment*" (Fred). An advantage of technology in this regard is that individuals feel more comfortable taking a break and walking away from the digital environment – whereas, in an office, they feel the pressure to stay at their desk: "*I usually like to have like 1 or 2 hours and then, you know, a couple hours to recover*" (Fred). This is made possible by the asynchronous capabilities of virtual environments.

## 5. Discussion

Our key contribution lies in three themes that illuminate how media capabilities can alleviate autistic people's experience of collaborative work environments. First, we identified media capabilities that can filter out distressing sensory, emotional, or cognitive input. For example, the capability to access a work environment remotely makes it possible to engage in meaningful aspects of work without suffering sensory distress from a physical office. Textual channels help reduce the emotionality of the received content, and other textual features, like email signatures, make it possible to automatically accompany text-based conversations with information about communication preferences. This helps reduce incoming flows of spontaneous interaction that can be disruptive. The digital capability to extract text from conversations (e.g., live captioning, transcripts) helps de-emphasize noisy audiovisual contexts and facilitates focused attention to relevant content. Similarly, camera-based communication downplays noisy input out-of-the-box by offering a zoomed-in view of an interlocutor's face. We refer to these capabilities as *digital filters*.

Another key theme pertains to the taxing nature of masking. Our findings align with previous research, indicating that masking drains mental health (cf. Hull et al., 2017), yet participants engage in it habitually. A significant burden involves distancing from one's authentic self, echoing existing literature (Pearson & Rose, 2021). Together with the digital filters, other media capabilities help curate an environment where individuals regain agency in how they relate to masking. Using a webcam as a proxy for eye contact is an example of using those media capabilities for assimilation goals (Hull et al., 2017) while alleviating the strain of masking. The visual frame available through audiovisual communication channels is another capability serving assimilation goals by concealing behaviors that may be considered unprofessional or socially unacceptable. This enables workers to fully engage with the interaction while being in control of what is visible in the frame, as the strain of monitoring everything else is alleviated. Our data also supports the connection thesis (Hull et al., 2017), as we found that participants leverage media capabilities that enable them to shape their way of presenting to build coworker relationships. The controlled visual frame is one such capability, another is the use of symbols to exert more control over the interpretation of their emotional displays or intentionality. For example, emojis clarify the tone that could otherwise be misinterpreted in audiovisual or plain text communication. Those capabilities help individuals can access a new level of relationship-building, as stigmatizing barriers are minimized or mediated. Media capabilities thus (i) provide assistive support that makes it less exhausting to mask and (ii) render masking unnecessary by concealing behaviors that pose threats to assimilation goals.  They effectively transform the work environment into a *digital stage* that inherently supports masking, easing its usual strain. In this space, autistic individuals redirect cognitive resources to less strenuous forms of masking.

The third theme describes the media capabilities that structure the temporal interplay between work demands and personal needs. As many workers may require additional time to process sensory, cognitive, and affective input, typical work routines often fail to accommodate those needs. Reprocessability and rehearsability were previously identified as important information-processing capabilities (Dennis et al., 2008). We note similar capabilities holding particular importance in neurodiverse collaboration because they act as a temporal buffer that reconciles individual needs with typical work demands. For example, capabilities that help craft content with greater awareness and intentionality, such as asynchronous channels or generative AI, support self-reflection before responding. These capabilities dynamically facilitate the better communication of one's authentic reaction or the more effective crafting of more suitable  responses. At the same time, they support connection goals as workers consciously leverage them to enhance their relationships with colleagues. The capability to momentarily disengage from ongoing interaction further makes room for recovery. As social demands can pile up and cause exhaustion, asynchronous channels make it possible for workers to momentarily leave the psychological space of work and engage in a psychological space of self-care. Slowing down and reflecting on the internal and external cues that are often more challenging to catch in typical work temporalities contributes to better performance, relationships, and individual well-being. Figure 1 illustrates the relationships between the three themes, conceptualizing what we call technorelief: digital collaboration media offer capabilities to filter out input from colleagues and the work environment (digital filters) as well as create a virtual space where autistic individuals are more empowered in how they present to others (digital stage) while creating an individual-scale time that is compatible with and enhances collaborative work (digital timeout).

**Table 2. Thematic analysis and definitions**

| Digital filters | Sensory input and cognitive overload | Uncontrollable characteristics of physical work environments (e.g., noise, temperature, movements, etc.) resulting in distraction, physical pain, and mental distress. |
|---|---|---|
| | Filtering effects of technology | Media capabilities to increase physical and psychological distance with socio-environmental triggers, creating a protective to preserve perceived homeostasis. |
| The digital stage | Masking to exhaustion | Self and socially inflicted pressure for typical behavior result in costly adaptation and camouflaging strategies. |
| | Digital space as a controlled stage | Media capabilities to structure a space where social performance is more easily controllable and less straining. |
| The digital timeout | Asynchronous socio-emotional processing | Informationally-rich situations and individual interoceptive challenges result in delayed processing of socio-emotional cues. |
| | Digital timeout for controlled reflection | Media capabilities that embed socio-emotional timeouts into the structure of the interaction for more effective self-reflection and presentation. |

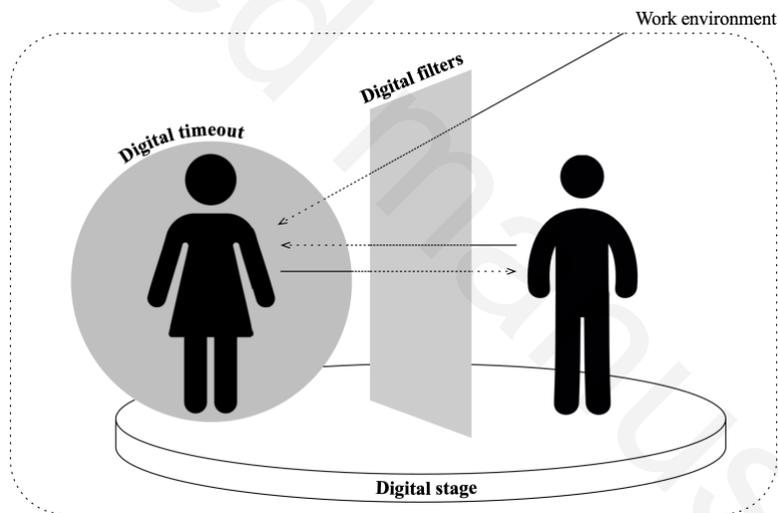

**Figure 1. The Technorelief Model.**

Our second most important contribution is the concept of *technorelief*. With it, we shed unprecedented light on the specific ways media capabilities interact with autistic adults' collaboration experiences. A large stream of research has studied the concept of technostress (Ayyagari et al., 2011), described as the anxiety, tension, or distress resulting from someone's inability to cope with the stressors created by technology. *Technorelief* describes the psychological relief achieved through the use of technology to cope with stressors from the work environment. We identified several media capabilities that drive technorelief through a network of sensory, cognitive, and affective mechanisms. We invite future research to further develop the concept, including in typically developing populations.

Third, our findings have implications for information-processing researchers. While there are extensive empirical studies of media richness, synchronicity, and naturalness theories, this work suggests that new analytical frameworks might be needed to include media capabilities that moderate sensory stimuli and affective cues.

## 6. Conclusion

This study explores the interaction of digital media with neurodiverse collaboration, focusing on the experiences of autistic individuals. By conducting a thematic analysis of 33 interviews, three key themes emerge. First, media capabilities act as *digital filters* that help manage overwhelming sensory inputs. Second, they aid in navigating the challenges of masking, allowing more agency in how individuals present themselves in the *digital stage*. Third, they enable a *digital timeout*, offering a temporal buffer for self-care and reflection. The concept of *technorelief* complements technostress, highlighting technology's positive role in managing collaboration stress for neurodivergent workers. This study offers insights into enhancing work experiences for neurodiverse individuals, fostering inclusive and supportive work environments.